\documentclass[graybox, nosecnum]{svmult} 

\usepackage{mathptmx}       
\usepackage{helvet}         
\usepackage{courier}        
\usepackage{type1cm}        
\usepackage{makeidx}         
\usepackage{graphicx}        
\usepackage{multicol}        
\usepackage{multirow}
\usepackage[bottom]{footmisc}
\usepackage[hidelinks]{hyperref}        
\usepackage{soul}            
\usepackage[sort&compress]{natbib}
\usepackage{amsmath,amssymb,mathtools}
\usepackage{doi}

\makeindex             

\begin{document}
\title*{Fast Flavor Transformations}
\author{Sherwood Richers \thanks{corresponding author} and Manibrata Sen}
\institute{Sherwood Richers \at Department of Physics, University of California Berkeley, Berkeley, CA 94720, \email{srichers@berkeley.edu}
\and Manibrata Sen \at Max-Planck-Institut für Kernphysik, Saupfercheckweg 1, 69117 Heidelberg, Germany, \email{manibrata@mpi-hd.mpg.de}}
%
%
\maketitle
\abstract{The neutrino fast flavor instability (FFI) can change neutrino flavor on time scales of nanoseconds and length scales of centimeters. It is expected to be ubiquitous in core-collapse supernovae and neutron star mergers, potentially modifying the neutrino signal we see, how matter is ejected from these explosions, and the types of heavy elements that form in the ejecta and enrich the universe. There has been a great deal of recent interest in understanding the role the FFI plays in supernovae and mergers, but the short length and time scales and the strong nonlinearity have prevented the FFI from being included consistently in these models. We review the theoretical nature of the FFI starting with the quantum kinetic equations, where the instability exists in neutron star mergers and supernovae, and how the instability behaves after saturation in simplified simulations. We review the proposed methods to test for instability in moment-based calculations where the full distribution is not available and describe the numerical methods used to simulate the instability directly. Finally, we close by outlining the trajectory toward realistic, self-consistent models that will allow a more complete understanding of the impact of the FFI in supernovae and mergers.}


\section{Introduction}
Core-collapse supernovae (CCSNe) and neutron star mergers (NSMs) are the only sites in the universe after the big bang where neutrinos are generated at sufficiently high densities that they are not only temporarily trapped by dense matter, but they interact and scatter with themselves in a way that drives a rich variety of strongly nonlinear effects. Prior to the start of the millenium, it was widely believed that the major channel of flavor conversions in a CCSN was due to resonant flavor conversions caused by the Mikheyev-Smirnov-Wolfenstein (MSW) effect \citep{wolfenstein_NeutrinoOscillationsMatter_1978,mikheyev_ResonantNeutrinoOscillations_1989, dighe_IdentifyingNeutrinoMass_2000}. However, this picture neglects the effect of the neutrino self-interactions, which was shown to be absolutely crucial for neutrino flavor evolution in a dense media \citep{kostelecky_SelfmaintainedCoherentOscillations_1995,samuel_BimodalCoherenceDense_1996,duan_CollectiveNeutrinoFlavor_2006,duan_SimulationCoherentNonlinear_2006,duan_CoherentDevelopmentNeutrino_2006}. These self-interactions manifest as a forward-scattering potential $\propto \sqrt{2}G_F n_\nu$ for a background neutrino density $n_\nu$ \citep{pantaleone_NeutrinoOscillationsHigh_1992}.
Neutral current interactions between neutrinos of different flavors, being flavor-blind, cause this forward-scattering potential to have off-diagonal components as well \citep{duan_CollectiveNeutrinoFlavor_2006}. As a result, a dense ensemble of neutrinos undergo collective flavor transformation, exhibiting a rich phenomenology.

Until the middle of the last decade, it was widely believed that these collective oscillations would lead to bipolar flavor conversions growing with a rate, roughly proportional to $\sqrt{(\Delta m^2) n_\nu}$, where $\Delta m^2$ is the neutrino mass splitting. Simple toy model analyses predicted these flavor conversions to take place close to the stalled shockwave at a radius of $\mathcal{O}(200)\,$km, thereby hinting at their possible role in aiding a shockwave-driven explosion. While most of the initial numerical studies were performed using a single-angle approximation, where neutrinos of all flavors were emitted with the same angle from a certain neutrinosphere, later studies showed that relaxing these approximations can lead to flavor synchronization, bipolar oscillations, spectral splits, and other instabilities, though at radii to large to affect the CCSN explosion mechanism \citep{fogli_CollectiveNeutrinoFlavor_2007,esteban-pretel_RoleDenseMatter_2008,sawyer_MultiangleInstabilityDense_2009,chakraborty_AnalysisMatterSuppression_2011,duan_SelfInducedSuppressionCollective_2011,dasgupta_RoleCollectiveNeutrino_2012}. 
We are still far from having a complete analytic picture of collective oscillations, but one can get an understanding of the onset of these instabilities using a linearized stability analysis \citep{banerjee_LinearizedFlavorstabilityAnalysis_2011,raffelt_AxialSymmetryBreaking_2013,chakraborty_CollectiveNeutrinoFlavor_2016}.

More recently, the different directional structures of neutrino distributions of different flavors has been recognized to cause an entirely different "fast" flavor instability (FFI), the subject of this review, that can occur in deep regions of CCSNe and NSMs not accessible to other flavor transformation mechanisms. Small perturbations can grow with a rate proportional to $n_\nu$ \citep{sawyer_SpeedupNeutrinoTransformations_2005} and can occur even for massless neutrinos, thereby being independent of the neutrino masses and mass hierarchy. These results were further confirmed in \cite{sawyer_MultiangleInstabilityDense_2009}, although they were derived using discretized neutrino angular distributions, which can give rise to spurious instabilities \citep{sarikas_SpuriousInstabilitiesMultiangle_2012}. The FFI was further substantiated by \cite{sawyer_NeutrinoCloudInstabilities_2016} by employing larger number of angular modes, in \cite{chakraborty_SelfinducedNeutrinoFlavor_2016} using simple toy models, and in \cite{dasgupta_FastNeutrinoFlavor_2017} realistic SN neutrino spectra. Since then, the FFI has inspired a great deal of work in order to understand the instability from an analytic point of view, simulate its nonlinear nature, discover where it is realized in nature, and determine its effects in astrophysical explosions.

This one of several recent reviews in the area of collective neutrino flavor transformations. \cite{duan_CollectiveNeutrinoOscillations_2010} reviews slow collective neutrino oscillation simulations and implications for supernovae. \cite{bellini_NeutrinoOscillations_2014} reviews neutrino oscillations applied to neutrino detectors, while \cite{mirizzi_SupernovaNeutrinosProduction_2016,horiuchi_WhatCanBe_2018} additionally review how detections of neutrinos from a supernova might inform astrophysics and fundamental physics. \cite{tamborra_NewDevelopmentsFlavor_2021} review a few phenomena associated with the FFI and \cite{capozzi_NeutrinoFlavorConversions_2022} review much of the recent work on the FFI and anticipated effects in CCSNe and NSMs. In this short review, we attempt a comprehensive overview of only recent developments related to the FFI, both on the numerical and analytical fronts, paying special attention to the methods used to find where the FFI occurs in nature and to probe the nature of the FFI once it takes hold. We do not discuss many-body effects \citep{pehlivan_InvariantsCollectiveNeutrino_2011,cervia_EntanglementCollectiveFlavor_2019, rrapaj_ExactSolutionMultiangle_2020, xiong_ManybodyEffectsCollective_2022,roggero_DynamicalPhaseTransitions_2021,patwardhan_SpectralSplitsEntanglement_2021,roggero_EntanglementManybodyEffects_2021,martin_ClassicalQuantumEvolution_2022,roggero_EntanglementCorrelationsFast_2022}, the exciting possibility of simulating neutrinos with quantum computers \citep{hall_SimulationCollectiveNeutrino_2021,yeter-aydeniz_CollectiveNeutrinoOscillations_2022,molewski_ScalableQubitRepresentations_2022,arguelles_NeutrinoOscillationsQuantum_2019}, non-standard interactions \citep{esteban-pretel_ProbingNonstandardNeutrino_2007,blennow_NonstandardNeutrinoneutrinoRefractive_2008,das_NewEffectsNonstandard_2017,dighe_NonstandardNeutrinoSelfinteractions_2018}, helicity coherence and sterile neutrinos \citep{volpe_ExtendedEvolutionEquations_2013,vaananen_LinearizingNeutrinoEvolution_2013,serreau_NeutrinoantineutrinoCorrelationsDense_2014,kartavtsev_NeutrinoPropagationMedia_2015,chatelain_HelicityCoherenceBinary_2017}, or wave packet separation \citep{giunti_CoherenceWavePackets_2004,akhmedov_CollectiveNeutrinoOscillations_2017,chatelain_NeutrinoDecoherencePresence_2020}. We hope this document helps to collect and structure work from the rapidly evolving field of neutrino fast flavor transformations, which bears relevance in so many areas of fundamental physics and astrophysics.

\section{Quantum Kinetic Equations}
A mean-field treatment of neutrino flavor conversions may be modeled using the occupation number matrix formalism, which can account for mixed states and possible loss of coherence due to collisions \citep{sigl_GeneralKineticDescription_1993,dolgov_NeutrinosEarlyUniverse_1981,raffelt_QuantumStatisticsParticle_1992,strack_GeneralizedBoltzmannFormalism_2005,giunti_FundamentalsNeutrinoPhysics_2007,vlasenko_NeutrinoQuantumKinetics_2014,volpe_NeutrinoQuantumKinetic_2015,blaschke_NeutrinoQuantumKinetic_2016}. The distribution of neutrinos is described by the Hermitian matrix-valued occupation number matrix $f^{ab}(\mathbf{x},\mathbf{p})$, where $a$ and $b$ are flavor indices and the spacetime position $\mathbf{x}$ and momentum $\mathbf{p}$ are four-vectors. Throughout this work, we assume that neutrinos are hyper-relativistic, so their momenta are null ($p^\alpha p_\alpha =-m^2\approx 0$). With this restriction, the distribution is a seven-dimensional quantity. Throughout this work we take care to indicate the structure of each quantity (matrix/tensor indices and spatial/momentum dependence) in the definition of each quantity, but suppress the additional markup elsewhere. The diagonal entries of this matrix are the occupation numbers for the corresponding neutrino species, while the off-diagonal elements encode quantum coherence between flavor states. Neglecting wave packet separation (valid on the short length/time scales of flavor instabilities; \citealt{akhmedov_CollectiveNeutrinoOscillations_2017}), the dynamics of the occupation number matrices are dictated by the quantum kinetic equation 
\begin{equation}
p^\alpha \frac{\partial f}{\partial x^\alpha} + \frac{d p^\alpha}{d\lambda}\frac{\partial f}{\partial p^\alpha}
= \epsilon \left({\mathcal C}- i \left[\mathcal{H}, f\right]\right)
\,\ .
\label{eq:eom_ch2}
\end{equation}
Here, $d\lambda$ is a differential unit of time in the frame comoving with the fluid and $p^\alpha \coloneqq dx^\alpha/d\lambda$. Factoring out the comoving-frame neutrino energy $\epsilon\coloneqq p^\alpha u_\alpha$ allows the part of the right hand side within the parentheses to be evaluated completely in a frame comoving with the background fluid defined by a four velocity $u^\alpha$. Throughout this work, we assume a $(+,-,-,-)$ metric convention, $a$ and $b$ are flavor indices, and $\alpha$, $\beta$, and $\gamma$ are spacetime indices. The Hamiltonian matrix $\mathcal{H}$ is composed of three terms, as
\begin{equation}
\mathcal{H}^{ab}(\mathbf{x}, \mathbf{p}) \coloneqq \mathcal{H}_\mathrm{vac}(\epsilon) + \mathcal{H}_\mathrm{matter}(\mathbf{x}) + \mathcal{H}_{\nu\nu}(\mathbf{x},\mathbf{v}) \,\,.
\label{eq:ham_ch2}
\end{equation}
$\mathcal{H}_\mathrm{vac}$ represents the contribution of the neutrino mass to its energy, which can be written in the flavor basis assuming hyper-relativistic neutrinos as
\begin{equation}
 \mathcal{H}_{{\rm vac}}^{ab}(\epsilon) \coloneqq U \frac{M^2}{2\epsilon}U^\dagger\,
\end{equation}
where $U^{ab}$ is the PMNS mixing matrix describing the rotation between the mass to flavor basis (see \cite{particledatagroup_ReviewParticlePhysics_2020} for values). $(M^2)^{ab}\coloneqq{\rm diag}\,(m_1^2,\,m_2^2,\,m_3^2)$ is the squared neutrino mass matrix. 

Neutrinos can interact with background matter (leptons, nucleons, other neutrinos) in a way that does not change the neutrino momentum (i.e., \textit{forward scattering}). The leading order interactions with background leptons and nucleons result in
\begin{equation}
    \mathcal{H}^{ab}_\mathrm{matter}(\mathbf{x})\coloneqq\sqrt{2}G_F \Lambda^{ab}(\mathbf{x}) \,\,,
\end{equation}
where $G_F$ is the Fermi coupling constant, $\Lambda^{ab}(\mathbf{x})=\delta^{ab}(n_a-n_{\bar{a}})$, and $n_a(\mathbf{x})$ are lepton number densities in the fluid rest frame. Note that in astrophysical environments where there are no muon or tauon neutrinos, an additional radiative correction can provide the leading order contribution distinguishing the $\nu_\mu$ and $\nu_\tau$ flavor states \citep{dighe_IdentifyingNeutrinoMass_2000}. The neutral current contributions from protons and electrons cancel. The contribution from neutrons is proportional to the identity matrix and as a result does not contribute to dynamics unless considering helicity or pair coherence. Because of this, they are usually neglected. 

The Hamiltonian contribution due to forward scattering off of other neutrinos (somewhat confusingly dubbed \textit{self-interaction}) is given by \citep{pantaleone_NeutrinoOscillationsHigh_1992}
\begin{equation}
\mathcal{H}_{\nu\nu}^{ab}(\mathbf{x},\mathbf{v}) \coloneqq \sqrt{2} G_F v^\alpha I_{1,\alpha}^{ab}(\mathbf{x}) \,\,,
\label{eq:self_ch2}
\end{equation}
where the angular moments of the neutrino lepton number density distribution are given by \begin{equation}
    \begin{aligned}
    I_0^{ab}(\mathbf{x}) &\coloneqq  \int \frac{d^2 \mathbf{v}}{4\pi} G(\mathbf{x},\mathbf{v}) \\
    I_{1,\alpha}^{ab}(\mathbf{x}) &\coloneqq \int \frac{d^2\mathbf{v}}{4\pi} G(\mathbf{x},\mathbf{v}) v_\alpha\\
    I_{2,\alpha\beta}^{ab}(\mathbf{x}) &\coloneqq \int \frac{d^2 \mathbf{v}}{4\pi} G(\mathbf{x},\mathbf{v}) v_\alpha v_\beta\\
    \end{aligned}
    \label{eq:moments}
\end{equation}
and so on. The null direction vector of the neutrino in an orthonormal tetrad comoving with the fluid is defined as $v^\beta = p^\alpha \hat{x}^{(\beta)}_\alpha/\epsilon$, where $\hat{x}^{(\beta)}_\alpha$ are the basis vectors defining the orthonormal tetrad. The differential neutrino lepton number distribution for each direction $\mathbf{v}$ is
\begin{equation}
    G^{ab}(\mathbf{x},\mathbf{v})\coloneqq\int \frac{\epsilon^2 d\epsilon}{2\pi^2}  \left[f(\mathbf{x},\epsilon,\mathbf{v})-\bar{f}^*(\mathbf{x},\epsilon,\mathbf{v})\right]\,\,,
    \label{eq:ELN}
\end{equation}

The evolution equations for antineutrinos is analogous; one must simply put bars over $f$, $\mathcal{H}$, and $\mathcal{C}$ in Equation~\ref{eq:eom_ch2}. In this case, $\bar{\mathcal{H}}=\bar{\mathcal{H}}_\mathrm{vac}+\bar{\mathcal{H}}_\mathrm{matter}+\bar{\mathcal{H}}_{\nu\nu}$, where $\bar{\mathcal{H}}_\mathrm{vac}=\mathcal{H}_\mathrm{vac}^*$, $\bar{\mathcal{H}}_\mathrm{matter}=-\mathcal{H}_\mathrm{matter}^*$, and $\bar{\mathcal{H}}_{\nu\nu}=-\mathcal{H}_{\nu\nu}^*$. Some works express the antineutrino evolution equations for a different quantity $\bar{f}'\coloneqq-\bar{f}^*$. This causes the integrand in Equations~\ref{eq:ELN} to be proportional to $(f+\bar{f}')$ and allows the antineutrino Hamiltonians to be written as  $\bar{\mathcal{H}}_\mathrm{vac}'=-\mathcal{H}_\mathrm{vac}$, $\bar{\mathcal{H}}_\mathrm{matter}'=\mathcal{H}_\mathrm{matter}$, and $\bar{\mathcal{H}}'_{\nu\nu}=\mathcal{H}_{\nu\nu}$.

Detailed collisional interaction rates for a single neutrino species have been developed and extensively used in CCSN and NSM simulations (e.g., \citealt{bruenn_StellarCoreCollapse_1985,burrows_NeutrinoOpacitiesNuclear_2006}). The two most simple interaction types can be generalized to matrix-valued QKE collision terms by defining an opacity matrix $\langle \kappa \rangle^{ab}=(\kappa_{\nu_a}+\kappa_{\nu_b})/2$, where $\kappa_{\nu_a}(\mathbf{x},\epsilon)$ are single-species opacities (and similarly for emissivity $\eta(\mathbf{x},\epsilon)$; \citealt{vlasenko_NeutrinoQuantumKinetics_2014,volpe_NeutrinoQuantumKinetic_2015,blaschke_NeutrinoQuantumKinetic_2016,richers_NeutrinoQuantumKinetics_2019}). Absorption and emission are modeled by
\begin{equation}
\mathcal{C}_\mathrm{abs/emit}^{ab}(\mathbf{x},\mathbf{p}) = \langle\eta\rangle^{ab} (\delta^{ab}-f^{ab}) - \langle\kappa\rangle^{ab}_\mathrm{abs}f^{ab}\,\,.
\end{equation}
Elastic, isotropic scattering is given by
\begin{equation}
\mathcal{C}_\mathrm{elastic\,\,scat}^{ab}(\mathbf{x},\mathbf{p}) = \langle\kappa\rangle^{ab}_\mathrm{scat} \int \frac{d^2\mathbf{v}'}{4\pi} \left[f^{ab}(\mathbf{\mathbf{x},\epsilon,v'}) - f^{ab}(\mathbf{x},\epsilon,\mathbf{v})\right]\,\,.
\end{equation}
A similar description of a more complete set of interactions is given in \cite{richers_NeutrinoQuantumKinetics_2019}.

Many calculations assume that there are only two flavors of neutrinos for simplicity, to turn each $3\times3$ matrix into a $2\times2$ matrix. This is often motivated by the fact that the distributions of $\nu_\mu$ and $\nu_\tau$ are very similar in CCSNe and NSMs. This reduces the computational cost of a calculation, makes the results visualizable using a Bloch vector (i.e., a point on the surface of a two-sphere), and in many cases produces qualitatively similar results as a three-flavor calculation. However, quantitative predictions of the net content of each flavor are not generally reliable and in certain cases phenomena arise with three flavors that do not occur with two flavors \citep{dasgupta_CollectiveThreeflavorOscillations_2008,dasgupta_SpectralSplitPrompt_2008,chakraborty_ThreeFlavorNeutrino_2020, capozzi_MuTauNeutrinosInfluencing_2020,shalgar_ThreeFlavorRevolution_2021, capozzi_NeutrinoFlavorConversions_2022}.

\section{The Fast Flavor Instability}
\label{sec:FFI}

Solving Eq.\,\ref{eq:eom_ch2} in its entirety for a realistic astrophysical environments is computationally intractable. However, it is possible to analytically solve for the evolution of small perturbations by linearizing the equations assuming all flavor-diagonal components are homogeneous. A flavor off-diagonal element of the occupation number matrix for neutrinos moving in direction $\mathbf{v}$ can be decomposed into plane wave solutions 
\begin{equation}
    f^{ab}(\mathbf{x},\mathbf{p}) = \frac{f^{aa}(\mathbf{p})-f^{bb}(\mathbf{p})}{2} Q^{ab}(\mathbf{v}) e^{-iK^\alpha x_\alpha}\,\,,
\end{equation}
with (real) amplitude $Q^{ab}(\mathbf{v})$ and (complex) four wave number $\mathbf{K}=(\omega,k)$. If we assume that a neutrino distribution is initially very nearly flavor diagonal (i.e., $Q\ll 1$), one can plug this into Equation~\ref{eq:eom_ch2}, keeping only terms linear in $Q$ and ensuring that $f^{ab}=f^{ba*}$. The resulting equation has solutions that satisfy \citep{banerjee_LinearizedFlavorstabilityAnalysis_2011, izaguirre_FastPairwiseConversion_2017, capozzi_FastFlavorConversions_2017, morinaga_LinearStabilityAnalysis_2018}
\begin{equation}
\mathrm{det}\left[\eta^{\alpha\beta} +\int \frac{d^2\mathbf{v}}{4\pi}\,G^{(ab)}(\mathbf{v}) \frac{v^{\alpha}v^{\beta}}{K'_\gamma v^\gamma}\right]=0 \,\, ,
\label{eq:polar}
\end{equation}
where $K_\alpha'\coloneqq K_\alpha - \sqrt{2} G_F (I_{1,\alpha}^{(ab)}+\Lambda^{(ab)}\hat{t}_\alpha)$ and $\hat{t}^\alpha=(1,0,0,0)$ is the timelike basis vector in the tetrad. We define the difference between two diagonal elements of a flavor matrix as
\begin{equation}
    A^{(ab)} \coloneqq A^{aa} - A^{bb}\,\,.
\end{equation}

Even though global simulations including neutrino quantum kinetics are not currently possible, one can probe the potential importance of the FFI in existing simulation by searching for crossings of the ELN. In data where the full distribution is available, one can do this explicitly. However, many calculations have limited data, usually in the form of moments defined in Equation~\ref{eq:moments}. In what follows, we will omit the flavor superscripts $(ab)$ with the understanding that an instability criterion applies to any pair of two flavors $a$ and $b$. When spacetime indices are omitted, the moment tensors are assumed to be contracted with a unit vector along the axis of symmetry. We label a method as "exact" if there is a one-to-one correspondence between the criterion and instability, "conservative" if it cannot indicate instability where it does not exist, and "approximate" if it is a best-guess procedure. Furthermore, we illustrate the diversity of models that have been searched for instability, along with the diversity of methods employed, in Table~\ref{tab:FFI_search}.

\underline{\textbf{Explicit (exact):}} One can integrate the distribution function to get the differential lepton number asymmetry. That is, there is a lepton number crossing and hence instability where $\mathrm{max}_{\mathbf{v}}(G^{(ab)})\mathrm{min}_{\mathbf{v}}(G^{(ab)})<0$.

\underline{\textbf{$\mathbf{k}_0$ (conservative)}}: \cite{dasgupta_SimpleMethodDiagnosing_2018} show that there is always some wavenumber $k=k_0$ that simplifies the dispersion relation. In this case, the dispersion relation takes the form $\mathrm{det} \left(\eta_{\alpha\beta} + I^{(ab)}_{2,\alpha\beta}/\omega\right)=0$. A complex value of $\omega$ satisfying the dispersion relation implies that the mode with wavenumber $k_0$ is unstable. In axisymmetric distributions, this instability criterion simplifies to $(I_0+I_2)^2-4(I_1)^2<0$.

\textbf{\underline{$\mathbf{\alpha}$ (conservative)}}: \cite{abbar_FastNeutrinoFlavor_2019,glas_FastNeutrinoFlavor_2020} propose that in regions of near-equilibrium (i.e. in the PNS or HMNS) and in directions opposite the LESA, it more practical to just look for locations that satisfy $\alpha=n_{\nu_e}/n_{\bar{\nu}_e}=1$ (equivalent to $I_0=0$), since if there are equal numbers of all neutrinos, crossings are inevitable.

\underline{\textbf{Polynomial (conservative)}}: \cite{abbar_SearchingFastNeutrino_2020} appeal to the presence of a crossing that is not restricted to $k=0$. Although the concept works for arbitrary distributions, it is most simple to express assuming an axisymmetric distribution. One can construct any linear combination of moments $I_\mathcal{F} = a_0 I_0 + a_1 I_1 + a_n I_n + ...$ such that the function $\mathcal{F}(\mu)=\sum a_n \mu^n$ is strictly positive for $\mu\in[-1,1]$. A crossing must exist and the distribution is thus unstable if $I_0I_\mathcal{F}<0$.

\underline{\textbf{Pendulum (conservative)}}: \cite{johns_FastFlavorInstabilities_2021} further indicate instability in axisymmetric distributions according to the \textit{resonant trajectory test} (unstable if $I_2^2 < I_1^2$) and the \textit{unstable pendulum test} (unstable if $I_2^2 \leq \frac{4}{5}I_1(5I_3-3I_1)$).

\underline{\textbf{Distribution Fit (approximate)}}: \cite{nagakura_ConstructingAngularDistributions_2021,nagakura_NewMethodDetecting_2021} propose a combination of a polynomial fits (fitted to a 1D CCSN simulation using $S_n$ transport) and a ray-tracing calculation to estimate the values of the radially ingoing and outgoing distributions. If $(f^{ee}-\bar{f}^{ee})_\mathrm{in} (f^{ee}-\bar{f}^{ee})_\mathrm{out}\leq 0$ and the heavy lepton neutrino distributions are equal to each other, there is a ELN crossing and thus instability.

\underline{\textbf{Maximum Entropy (approximate)}}:Assuming the neutrino and antineutrino distributions follow the \textit{maximum entropy} angular distribution of of \cite{cernohorsky_MaximumEntropyDistribution_1994} (i.e., $f(\mu)\sim \exp(\mu Z)$, where $\mu$ is the cosine of the angle from the flux direction), one can obtain the parameter $Z$ for each distribution to give it the appropriate net flux. The presence of crossings can be determined analytically from these assumed distributions \citep{richers_EvaluatingApproximateFlavor_2022}.

\section{Direct Simulation}

\begin{figure}
    \centering
    \includegraphics[width=\linewidth]{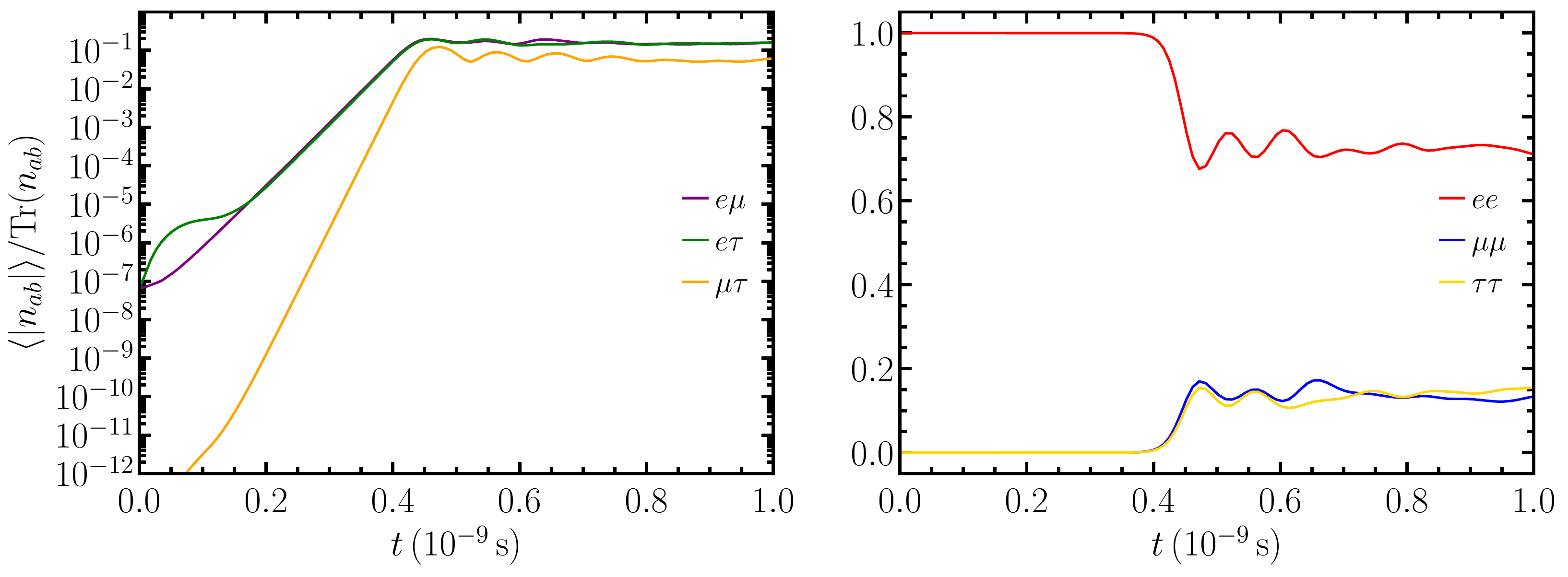}
    \caption{An example simulations of the FFI using {\tt Emu}. The initial distributions are defined by $n_{\nu_e}=4.89\times10^{32}\,\mathrm{cm}^{-3}$, $n_{\bar{\nu}_e}=0.5 n_{\nu_e}$, and $n_{\nu_\mu}=n_{\bar{\nu}_\mu}=n_{\nu_\tau}=n_{\bar{\nu}_\tau}=0$. Both the $\nu_e$ and $\bar{\nu}_e$ distributions decrease linearly with the cosine of the angle from opposite ends of the $z$ axis, with flux factors of $1/3$. Left: exponential growth of the flavor off-diagonal components of the neutrino number density and saturation of the instability at $t\approx0.4\,\mathrm{ns}$. Right: Evolution of the flavor diagonal components of the number density. The final abundance of each neutrino flavor depends on the particular initial conditions and only results in complete mixing in special cases.}
    \label{fig:FFI_Simulation}
\end{figure}

While linear stability analysis indicates that small perturbations will grow exponentially, what happens when the perturbations are no longer small? There is some analytical work predicting the nonlinear evolution of the FFI in homogeneous/semi-isotropic cases \citep{padilla-gay_FastFlavorConversion_2021} and crude estimates of post-instability flavor content \citep{bhattacharyya_FastFlavorDepolarization_2021,bhattacharyya_ElaboratingUltimateFate_2022}, but numerical simulations are still required to understand more general cases and to validate approximations made in analytical studies. Here we briefly review the numerical techniques that have been used to study the fast flavor instability. Differences fall largely into two categories: discretization scheme and assumed symmetries. \cite{richers_CodeComparisonFast_2022} compare several different methods and show that each has advantages and weaknesses. We neglect a discussion of methods used to treat other phenomena, such as slow collective oscillations, the MNR, and the multi-azimuthal-angle instability.

\textbf{\underline{Discretization Scheme:}}
Although the results of numerically converged simulations should not depend on the discretization scheme, it is important to understand the approximations and limitations involved. The majority of simulations employ the $\mathbf{S_n}$ scheme, in which phase space is divided into discrete blocks, each of which contains some amount of radiation. One then uses discrete derivatives to evaluate the advection term to determine how neutrinos move from block to block (though there are still many ways to do this in detail; \citealt{wu_CollectiveFastNeutrino_2021}). The approximation lies in the finite size of the blocks. \cite{zaizen_NonlinearEvolutionFast_2021} instead discretize space into \textbf{Fourier modes}, leaving momentum space discretized as in the $S_n$ scheme. This makes the spatial derivatives in the advection term easier to evaluate, but the advection term then couples every Fourier mode to every other Fourier mode, which can become computationally expensive for large systems. The approximation lies in the finite size of the blocks in momentum-space and the finite number of spatial Fourier modes. The \textbf{Particle in cell} method discretizes the radiation field into particles \citep{richers_ParticleincellSimulationNeutrino_2021}. This makes evaluation of the advection term more simple (the particles simply move in straight lines), but the Hamiltonian becomes more difficult to treat because one has to interpolate the full radiation field to the location of every particle. The approximation lies in the finite number of particles and the finite number of spatial grid cells that are used to evaluate the Hamiltonian. Finally, \cite{johns_FastOscillationsCollisionless_2020} expand momentum space into \textbf{angular moments}. This causes the evolution equation for each moment to depend on higher moments, such that one must approximate the system by cutting of the tower of equations at some moment order by applying a closure relation. Although it has yet seen use in the FFI literature, one can also discretize momentum space in terms of \textbf{spherical harmonics} (e.g., \citealt{duan_FlavorOscillationModes_2013}), which is conceptually similar to a moment expansion.

\textbf{\underline{Symmetries:}} It is common to simplify calculations by assuming homogeneity in one or more directions. We use 0D, 1D, 2D, and 3D to describe simulations that allow for inhomogeneity in no, one, two, or three directions. Allowing for inhomogeneity permits the presence of modes with nonzero wavenumber in that direction. The FFI is fundamentally a multi-direction phenomena, so all FFI simulations allow for some degree of anisotropy. Angular distributions generally assume one of several common symmetries. \textbf{Beam models} assume that neutrinos move along a small number of discrete beams (\textbf{0D}: \citealt{sawyer_ClassicalInstabilitiesQuantum_2004,sawyer_SpeedupNeutrinoTransformations_2005,sawyer_MultiangleInstabilityDense_2009,dasgupta_FastNeutrinoFlavor_2018,hansen_EnhancementDampingFast_2022}). Minimally complex models that allows for both inhomogeneity and anisotropy in one Cartesian direction are \textbf{planar geometry} and the \textbf{neutrino line model}, where neutrinos are allowed only to travel with velocity components specified by a single angle coordinate (\textbf{0D}:\citealt{abbar_FastNeutrinoFlavor_2019}, \textbf{1D}:\citealt{duan_FlavorInstabilitiesNeutrino_2015a,mirizzi_SelfinducedFlavorInstabilities_2015}, \textbf{2D}:\citealt{shalgar_NeutrinoPropagationHinders_2020,padilla-gay_MultiDimensionalSolutionFast_2021}). In order to discuss neutrino distributions directly relevant to approximately spherical core-collapse supernovae, the majority of recent simulations assume \textbf{axial symmetry} around the radial direction (\textbf{0D}: \citealt{dasgupta_FastNeutrinoFlavor_2017,dasgupta_FastNeutrinoFlavor_2018,johns_FastOscillationsCollisionless_2020,kato_NeutrinoTransportMonte_2021,padilla-gay_NeutrinoFlavorPendulum_2022,padilla-gay_FastFlavorConversion_2021,sasaki_DynamicsFastNeutrino_2021,shalgar_ChangeDirectionPairwise_2021,shalgar_ThreeFlavorRevolution_2021,xiong_StationarySolutionsFast_2021,johns_CollisionalFlavorInstabilities_2021}, \textbf{1D}: \citealt{martin_DynamicFastFlavor_2020,abbar_SuppressionScatteringInducedFast_2021,bhattacharyya_FastFlavorDepolarization_2021,duan_FlavorIsospinWaves_2021,martin_FastFlavorOscillations_2021,wu_CollectiveFastNeutrino_2021,zaizen_NonlinearEvolutionFast_2021,sigl_SimulationsFastNeutrino_2022,abbar_SuppressionFastNeutrino_2022,capozzi_SupernovaFastFlavor_2022,bhattacharyya_ElaboratingUltimateFate_2022,nagakura_TimedependentQuasisteadyGlobal_2022,george_COSENuCollective_2022}). Finally, one can allow for \textbf{general anisotropy} (\textbf{0D}:\citealt{shalgar_DispellingMythDense_2021,shalgar_SymmetryBreakingInduced_2021}, \textbf{1D}:\citealt{zaizen_ThreeflavorCollectiveNeutrino_2021,richers_ParticleincellSimulationNeutrino_2021}, \textbf{2D}:\citealt{bhattacharyya_FastFlavorOscillations_2021}, \textbf{3D}:\citealt{richers_NeutrinoFastFlavor_2021}). 

\textbf{\underline{Other Considerations:}} First, the vast majority of simulations assume flavor transformation only between two flavors, and relatively few consider \textbf{three flavors} \citep{capozzi_MuTauNeutrinosInfluencing_2020,richers_ParticleincellSimulationNeutrino_2021,shalgar_SymmetryBreakingInduced_2021,shalgar_ThreeFlavorRevolution_2021,zaizen_NonlinearEvolutionFast_2021,capozzi_SupernovaFastFlavor_2022}. Second, even when simulations allow for inhomogeneity, they generally assume periodic boundary conditions. This allows modes with $k\neq0$ to grow, but does not allow for \textbf{gradients} associated with dynamics on larger scales \citep{sigl_SimulationsFastNeutrino_2022}. Third, the interpretation of a simulation can be muddied by the form of the perturbations imposed in the \textbf{initial conditions} \citep{martin_DynamicFastFlavor_2020,wu_CollectiveFastNeutrino_2021}. Fourth, a full suite of \textbf{collisional interactions} has only been simulated assuming both isotropy and homogeneity, such that the FFI cannot arise \citep{richers_NeutrinoQuantumKinetics_2019}. \cite{johns_CollisionalFlavorInstabilities_2021} use physically motivated initial conditions and interaction rates in a homogeneous two-moment calculation to demonstrate interactions between collisional instabilities and the FFI.  First steps are also being taken to include toy-model elastic neutral-current scattering effects in simulations of the FFI \citep{martin_FastFlavorOscillations_2021,sasaki_DynamicsFastNeutrino_2021,shalgar_ChangeDirectionPairwise_2021,hansen_EnhancementDampingFast_2022}. They show that the isotropizing effect of scattering can keep distributions in an unstable state for a longer period of time, thus enhancing net flavor transformation, but if the collisions are too strong flavor transformation can be hindered. Finally, determining whether quantum many-body correlations (i.e., beyond the mean field limit) are important with macroscopic numbers of neutrinos, \cite{roggero_EntanglementCorrelationsFast_2022} suggests that the emergence of significant many-body correlations in distributions unstable to the FFI depends on the particular distribution.

\section{Core-Collapse Supernovae and Neutron Star Mergers}

\begin{table}
\centering
\begin{tabular}{llcccc}
    Progenitor ($M_\odot$) & Transport & Search Method & Reference Search \\ \hline 
    $11.2$  & 1D 3F VET  & Explicit & \cite{tamborra_FlavordependentNeutrinoAngular_2017}\\
    $25$  & 1D 3F VET  & Explicit & \cite{tamborra_FlavordependentNeutrinoAngular_2017}\\ 
   $15$  & 1D 3F VET  & Exp., Poly. & \cite{capozzi_FastNeutrinoFlavor_2021}\\
    $18.6$  & 1D 6F VET & Polynomial & \cite{capozzi_FastNeutrinoFlavor_2021}\\ 
    $20$  & 1D 6F VET & Polynomial & \cite{capozzi_FastNeutrinoFlavor_2021}\\ 
    $11.2$  & 2D 3F Snapshot $S_n$ & Explicit & \cite{abbar_FastNeutrinoFlavor_2020}& \\ 
    $11.2$ & 3D 3F Snapshot $S_n$ & Explicit & \cite{abbar_FastNeutrinoFlavor_2020}& \\ 
    $27$  & 3D 3F Snapshot $S_n$ & Explicit & \cite{abbar_FastNeutrinoFlavor_2020}\\ 
    $11.2$  & 1D 3F $S_n$ & Explicit & \cite{morinaga_FastNeutrinoflavorConversion_2020} \\ 
    $11.2$  & 2D 3F $S_n$  & Explicit & \cite{m.d.azari_LinearAnalysisFastpairwise_2019,m.d.azari_FastCollectiveNeutrino_2020}\\ 
    $11.2$  & 1D 3F $S_n$ & Explicit, Fit & \cite{nagakura_NewMethodDetecting_2021} \\ 
    $11.2$  & 2D 3F $S_n$ & Explicit, Fit &  \cite{nagakura_NewMethodDetecting_2021}\\ 
    $15$ (Rotating)  & 2D 3F $S_n$  & Explicit & \cite{harada_ProspectsFastFlavor_2022}\\ 
    $9$  & 3D 3F M1 & $k_0$, $\alpha$, Poly. & \cite{abbar_CharacteristicsFastNeutrino_2021} \\ 
    $20$  & 3D 3F M1  & $k_0$, $\alpha$, Poly. & \cite{abbar_CharacteristicsFastNeutrino_2021} \\ 
    $9-25$  & 3D 3F M1 & Fit & \cite{nagakura_WhereWhenWhy_2021}\\ 
    \hline
    $0.3 + 3$ BHD & 2D 2F Surface RT & Explicit & \cite{wu_ImprintsNeutrinopairFlavor_2017a} \\ 
    $1.35 + 1.35$ NSM & 2D 2F Surface RT & Explicit & \cite{george_FastNeutrinoFlavor_2020} \\ 
    $1.25 + 1.45$ NSM & 2D 2F Surface RT & Explicit & \cite{george_FastNeutrinoFlavor_2020} \\ 
    $0.07 + 3$ BHD & 3D 3F M1 & $k_0$ & \cite{li_NeutrinoFastFlavor_2021} \\
    $0.3 + 3$ BHD & 2D 2F M1 & Poly. & \cite{just_FastNeutrinoConversion_2022}
\end{tabular}
\caption{List of CCSN (above horizontal line) and NSM (below horizontal line) simulations that have been analyzed for crossings, along with a few details of the analysis method. The first column shows the mass of the CCSN progenitor, the component masses of the neutron stars in a NSM, or the disk and black hole mass of a black hole disk (BHD) system that does not simulate the merger itself. Both BHD calculations employ a black hole with a spin parameter of 0.8. The second column shows the transport method used to look for crossings (not necessarily that used in the simulation). 1D, 2D, and 3D indicate the number of spatial dimensions not approximated by a symmetry, and [2-6]F indicates the number of neutrino flavors involved in the calculation. $S_n$ indicates the neutrino field is defined along many discrete rays. Snapshot $S_n$ implies that the radiation field was post-processed on a simulation snapshot rather than evolved real-time in the simulation. VET means variable Eddington tensor, which also includes the neutrino field along discrete rays. M1 indicates a two-moment system with analytic closure. Surface RT means the neutrino distributions at a particular location are determined by tracing rays to a hard emission surface calculated approximately from the fluid data. The third column describes the search method, all of which are described in the main text. The final column shows the reference(s) with the most recent or complete crossing analysis. }
\label{tab:FFI_search}
\end{table}

Given the microscopic nature of the FFI, the question remains: where does the FFI occur in nature and what does it do? Although there are many reactions that contribute to the neutrino distribution in CCSNe and NSMs \citep{bruenn_StellarCoreCollapse_1985,burrows_NeutrinoOpacitiesNuclear_2006}, the charged-current absorption/emission reaction causes the distributions of electron neutrinos and antineutrinos to be significantly different from other species:
\begin{equation}
    p + e^- \leftrightarrow n + \nu_e\,\,.
\end{equation}
How this and other interactions cause the distributions (and instabilities) to manifest depends on a large number of factors, including turbulent relativistic hydrodynamics, the properties of matter above nuclear densities, the astrophysics creating the initial conditions of the event, and non-equilibrium neutrino radiation transport. Based on state of the art simulations of CCSNe and NSMs, the FFI seems to be robustly present, but the astrophysical implications are only beginning to be explored.

\textbf{\underline{PNS Convection}}: Crossings are found in the PNS convection region in all multidimensional models. Within the protoneutron star, the neutrinos are trapped and the above reaction is approximately in equilibrium. The electron neutrino chemical potential is then $\mu_{\nu_e}=\mu_p+\mu_e-\mu_n$. From a permutation of that reaction, $\mu_{\bar{\nu}_e}=-\mu_{\nu_e}$. In the region between $10\,\mathrm{km}\lesssim r \lesssim 20\,\mathrm{km}$ for the first few hundred milliseconds after core bounce, the electron fraction is relatively low, so the large $\mu_n$ can approximately cancel $\mu_e$ to make $\mu_{\nu_e}\approx \mu_{\bar{\nu}_e}\approx 0$. That is, there are a similar number of electron neutrinos and antineutrinos, so the small anisotropies caused by PNS convection and slightly different opacities between $\nu_e$ and $\bar{\nu}_e$ can induce crossings \citep{abbar_OccurrenceFastNeutrino_2019}. However, given that the chemical potential of all of the other flavors is also $\mu_{\nu_\mu}\approx\mu_{\nu_\tau}\approx\mu_{\bar{\nu}_\mu}\approx\mu_{\bar{\nu}_\tau}\approx 0$. All neutrino flavors have approximately the same distributions anyway, and it is unclear if flavor mixing will have any significant effect on the dynamics or the neutrino signal \citep{glas_FastNeutrinoFlavor_2020}.

\textbf{\underline{Under the Shock}}: Whether there are crossings in this region appears to depend on the details of the radiation transport method \citep{m.d.azari_FastCollectiveNeutrino_2020,nagakura_FastpairwiseCollectiveNeutrino_2019}. Exploding models seem to have more unstable regions \citep{abbar_CharacteristicsFastNeutrino_2021,nagakura_WhereWhenWhy_2021,capozzi_FastNeutrinoFlavor_2021}, stellar rotation may help suppress instability \citep{harada_ProspectsFastFlavor_2022}, the presence of multidimensional effects encourages crossings \citep{nagakura_NewMethodDetecting_2021} (though crossings can appear in 1D simulations; \citealt{capozzi_FastNeutrinoFlavor_2021}), and the LESA seems to guarantee crossings in some regions \citep{abbar_SearchingFastNeutrino_2020,abbar_CharacteristicsFastNeutrino_2021}.

\textbf{\underline{Above the Shock}}: Crossings are found outside the shock in all models. Infalling stellar material above the shock is rich with large nuclei before they dissociate upon passing through the shock front. Neutrinos at energies present in supernovae interact coherently with all of the nucleons in the nucleus, such that the cross section scales roughly as the square of the number of nuclei and the square of the neutrino energy. The antineutrinos that escape from a supernova have a higher average energy, and so scatter more efficiently from these nuclei. Although there are generally more electron neutrinos than electron antineutrinos moving outward (a consequence of the dense matter becoming more neutron rich), it turns out that there are more ingoing (i.e. scattered) electron antineutrinos than electron neutrinos \citep{morinaga_FastNeutrinoflavorConversion_2020}. Although there are very few ingoing neutrinos of any type, this technically constitutes a ELN crossing. Given that this crossing is very small, it is yet unclear if the resulting FFI can cause significant flavor change \citep{bhattacharyya_FastFlavorDepolarization_2021,richers_ParticleincellSimulationNeutrino_2021,wu_CollectiveFastNeutrino_2021,abbar_SuppressionFastNeutrino_2022}. However, independently from the FFI, this "halo" of scattered neutrinos could significantly modify the distribution of neutrinos observable on Earth \citep{cherry_NeutrinoScatteringFlavor_2012,sarikas_SupernovaNeutrinoHalo_2012,cherry_HaloModificationSupernova_2013,sawyer_NeutrinoCloudInstabilities_2016,cirigliano_CollectiveNeutrinoOscillations_2018,cherry_TimeFlightSupernova_2020,zaizen_NeutrinoHaloEffect_2020}. Note that \cite{capozzi_FastNeutrinoFlavor_2021} show that these crossings are not actually present at mu=-1, but that there is a double crossing in the middle. \cite{nagakura_TimedependentQuasisteadyGlobal_2022} perform the first large-scale models of the FFI above the shock using imposed boundary conditions.

\textbf{\underline{PNS cooling:}} \cite{xiong_PotentialImpactFast_2020} indicate that, should the FFI occur in the cooling phase of a PNS, it could significantly increase mass loss rates and affect nucleosynthesis with more proton-rich conditions.

\textbf{\underline{Neutron Star Mergers}}
\cite{wu_FastNeutrinoConversions_2017,wu_ImprintsNeutrinopairFlavor_2017a,george_FastNeutrinoFlavor_2020} find unstable regions in a NSM disk by determining the emission surface of NSM accretion disk and ray-tracing to estimate full neutrino distributions. They suggest that the flavor transformation is quite ubiquitous, that there is enhancement of r-process element production for a disk around a black hole, and that there is little impact on the r-process for a disk around a hypermassive neutron star. \cite{li_NeutrinoFastFlavor_2021} simulate a neutron star merger disk around a black hole assuming flavor equipartition wherever there is instability according to the $k_0$ test, which results in enhanced production of r-process elements. \cite{just_FastNeutrinoConversion_2022} also simulate a merger disk, but assume instability above a critical flux factor. They vary the flavor transformation prescription, disk mass, and MHD treatment, also finding moderately enhanced r-process yields in most cases. \cite{padilla-gay_MultiDimensionalSolutionFast_2021} simulate flavor transformation a toy model of a NSM disk that suggests minimal net flavor transformation even in the presence of instabilities. More work is needed to include the effects of the FFI self-consistently and in conjunction with collisions.

\section{Future Directions}
There has been a great deal of progress since the FFI was proposed and discovered in models of CCSNe and NSMs, but there is still a lot of work to be done before its potential astrophysical effects are sorted out. There are a number of significant open questions: What is the effect of combining flavor transformation and collisional processes in a realistic CCSN or NSM? Can instabilities in disparate locations affect each other? Are effects in simplified simulations as strong when combined with the immense complexity of physical processes in multidimensional astrophysical explosions? How can flavor transformation be reliably incorporated into global CCSN and NSM simulations?

One approach is to model the FFI as a small-scale phenomenon using a surrogate or sub-grid model. This work has already begun, and enable known effects to be included in large-scale simulations, but precludes finding new effects from the underlying instability operating and propagating on larger scales. Another approach is to simplify, truncate, or approximate the quantum kinetic equations in a way that is able to maintain the important features of the instability while ignoring unimportant features of the solution. The moment decomposition of the quantum kinetic equations is one example of this approach that shows promise for enabling the inclusion of flavor transformation effects on much larger scales and with a greater amount of additional physics than would be possible with a direct approach. Alternatively, one could artificially reduce the separation of scales between collisional and flavor transformation processes by effectively decreasing the interaction potential and extrapolating back to the full strength \citep{nagakura_GRQKNTCodeGeneralRelativistic_2022}, though this extrapolation must be carefully checked to avoid large systematic errors. Finally, all of this work assumes a mean-field treatment of the neutrino quantum states, when in fact neutrinos will be entangled with other neutrinos. Increasingly realistic calculations of many-body neutrino flavor transformation (potentially requiring quantum computers) will help inform whether many-body effects manifest in supernova and merger conditions. Although many of these questions seem solvable within the next decade, neutrino physics never ceases to yield new surprises, and the FFI is but one of many exciting effects in the incredibly complex environments in supernovae and mergers.

\bibliographystyle{apsrev4-2}
\bibliography{references.bib}

\end{document}